\documentclass[sigconf, screen]{acmart}

\AtBeginDocument{%
  }

\usepackage{xcolor}
\usepackage[skip=5pt]{caption} 

\usepackage{enumitem}
\setlist{nolistsep}

\setcopyright{none} 
\copyrightyear{2025}
\acmYear{2025}
\acmDOI{XXXXXXX.XXXXXXX}

\acmConference[SC25]{SC25}{Nov. 16--21, 2025}{St. Louis, USA}
\acmISBN{000-0-0000-0000-0/00/00}

\usepackage{caption}
\newcommand{\Subitem}[1]{\begin{itemize}
    \item[-] #1\end{itemize}}

\usepackage{color}
\usepackage[most]{tcolorbox}
\newcommand{\lan}[1]{\textcolor{blue}{zlan:#1}}

\newtcolorbox{mybox}{
enhanced,
boxrule=0pt,frame hidden,
borderline west={4pt}{0pt}{green!75!black},
colback=green!10!white,
sharp corners
}

\newtcolorbox{oppbox}{
enhanced,
boxrule=0pt,frame hidden,
borderline west={4pt}{0pt}{violet!75!black},
colback=violet!10!white,
sharp corners
}

\begin{document}

\title{Extracting Practical, Actionable Energy Insights from Supercomputer Telemetry and Logs}


\author{Melanie Cornelius}
\affiliation{%
   \institution{University of Illinois at Chicago}
   \city{Chicago}
   \state{IL}
   \country{USA}}
\email{mcornelius@anl.gov, mcorn18@uic.edu}

\author{Greg Cross}
\affiliation{%
   \institution{University of Illinois at Chicago}
   \city{Chicago}
   \state{IL}
   \country{USA}}
\email{grog@uic.edu}

\author{Shilpika}
\affiliation{%
   \institution{Argonne National Laboratory}
   \city{Lemont}
   \state{IL}
   \country{USA}}
\email{shilpika@anl.gov}

\author{Matthew T. Dearing}
\affiliation{%
   \institution{University of Illinois at Chicago}
   \city{Chicago}
   \state{IL}
   \country{USA}}
\email{mdear2@uic.edu}

\author{Zhiling Lan}
\affiliation{%
   \institution{University of Illinois at Chicago}
   \city{Chicago}
   \state{IL}
   \country{USA}}
\email{zlan@uic.edu}

\begin{abstract}
As supercomputers grow in size and complexity, power efficiency has become a critical challenge, particularly in understanding GPU power consumption within modern HPC workloads.
This work addresses this challenge by presenting a data co-analysis approach using system data collected from the Polaris supercomputer at Argonne National Laboratory.
We focus on GPU utilization and power demands, navigating the complexities of large-scale, heterogeneous datasets.
Our approach, which incorporates data preprocessing, post-processing, and statistical methods, condenses the data volume by 94\% while preserving essential insights.
Through this analysis, we uncover key opportunities for power optimization, such as reducing high idle power costs, applying power strategies at the job-level, and aligning GPU power allocation with workload demands.
Our findings provide actionable insights for energy-efficient computing and offer a practical, reproducible approach for applying existing research to optimize system performance.

\end{abstract}

\keywords{System logs, power efficiency, GPU usage, workload characteristics}


\maketitle

\section{Introduction}
As high-performance computing (HPC) rapidly evolves to meet the growing demands of scientific research, supercomputers grow in size and complexity.
Today, these systems are pushing the boundaries of computational capability with an increasing reliance on heterogeneity: nine out of the top ten supercomputers in the world feature heterogeneous CPU-GPU architectures \cite{top500}.
While extraordinary in their capabilities, these systems come at a high cost --- particularly in power consumption, with GPUs accounting for a substantial portion of the overall energy demand \cite{Abe2012GPU}.

We aim to leverage system-level data to gain deeper insights into GPU power consumption on supercomputers, with the ultimate goal of identifying opportunities for system-wide GPU energy optimization. 
Modern supercomputers continuously generate extensive log and monitoring data that capture telemetry, system states, and job scheduling details. 
Throughout this study, this data is collectively referred to as \textit{system logs}.
While the volume and heterogeneity of this data—arising from disparate sources, formats, schemas, and collection frequencies—pose significant challenges for analysis, they also present a unique opportunity: once curated and integrated, this data 
offers a critical lens through which we can observe system behavior and performance under real-world workloads.
This opens the door to actionable insights in resource efficiency \cite{9622850, 9556013}.

Historically, system log analysis has supported a wide range of research goals. Characterizing workload behavior is a common research objective, often involving analysis of compute, network, I/O, and memory \cite{DT-sc20, iccs2018, jsspp17, 9912707, 1526010,DT-ipdps20}. As GPUs have become central to high-performance computing, studies have increasingly focused on GPU utilization. More recently, power consumption in relation to GPUs has received growing attention, particularly through system-level case studies \cite{10820672, nersc23}.
However, relatively little attention has been given to using broad system data to uncover practical opportunities for applying power savings strategies on GPUs.
By carefully refining the data and allowing it to guide our exploration, we identify areas where established techniques may be effectively applied to optimize power efficiency.

In this work, we investigate the relationship between GPU workloads and power consumption using data from a production HPC environment. 
We analyze a year's worth of job scheduler event logs and GPU telemetry metrics from Polaris (ranked 47th in the Top500 list \cite{top500}).
Our focus is on uncovering meaningful insights into how workload behavior influences energy consumption, with the goal of identifying opportunities where power-saving strategies can be effectively applied to this system.
We also address handling these logs.
While data analysis has historically been common for HPC system logs \cite{DT-sc20, tuncer2017big, 10763585, Li2023},  the detailed analysis required to identify GPU power efficiency opportunities remains underdiscussed in the literature.
Ultimately our goal is \textbf{to offer a practical and reproducible approach that identifies actionable insights and contributes to the broader conversation on energy-aware computing}.

Specifically, this SOP study makes the following contributions:

\begin{itemize} 
    \item {Develops a systematic strategy, accompanied by the release of our dataset and open-source analysis tool, for co-analyzing diverse system logs. This provides a repeatable process to condense and tailor data for power-focused workload analysis(\S \ref{data-section}).}
    \item {Uncovers new insights and lessons for energy efficient computing, and proposes feasible opportunities for practical power optimization. These include:}
    \Subitem{Revisiting GPU power provisioning and reducing idle power, e.g., by enabling software-controlled P-states, can lead to substantial power gains (\S 4).}
    \Subitem{Job-level power management that aligns GPU power allocation with workload demands presents an effective approach for optimizing power efficiency (\S 5).}
    \Subitem{Strategies that leverage the bursty and inconsistent GPU memory access patterns can further enhance power efficiency (\S 6).}
    \Subitem{GPU memory allocation itself does not impact GPU power usage; instead, optimizing memory utilization is key (\S 7).}
\end{itemize}

\section{The Polaris System}
This work examines one year of system log data from \emph{Polaris}, a production supercomputer at Argonne National Laboratory. It is ranked 47th on the TOP500 list as of November 2024 \cite{top500}.


\subsection{System Overview}\label{system}
Polaris's compute capacity is 560 nodes, composed of 280 ProLiant XL645d Gen10 Plus dual-node systems. Each node contains a 32-core 2.8 GHz AMD Epyc 7543P ``Milan'' CPU, 512 GB system RAM, and four NVIDIA A100 SXM4 40GB GPUs. Nodes communicate over an HPE Slingshot 11 fabric, with each node connected by dual 200 Gb/s Cassini-based NICs. The fabric is a 40-group dragonfly topology with 14 nodes per group. 

The NVIDIA A100 SXM4 40 GB is an ``Ampere''-series GPU with 6,912 shading units and 432 tensor cores with a clock rate of 1.095 GHz base and 1.410 GHz boost. The 40 GB enhanced second-generation high bandwidth memory (\textit{HBM2e}) can transfer 1.56 TB/s using a clock rate of 1.215 GHz. The four GPUs in each Polaris node communicate directly over an unswitched NVLink interconnect on the NVIDIA ``Redstone'' board. Each GPU has a TDP (thermal design power, which describes theoretical max power draw) of 400 W and idles at just over 50 W in its maximum performance state. 

The maximum power draw for a Polaris node is approximately 2200 W, measured directly from the system's PDUs. 
Of this, the four A100 GPUs consume up to 1600 W, accounting for more than \textbf{half} of the total node power. 
Clearly, reducing GPU power on Polaris would significantly impact the overall node power draw.

The Polaris supercomputer runs on the SUSE Linux Enterprise Server (SLES) release 15 operating system and is managed by the HPE Performance Cluster Manager (HPCM). Jobs are batch-scheduled through Altair PBS Pro and launched by the HPE Parallel Application Launch Service (PALS). Table \ref{queue-table} describes the job queues specified on Polaris.

\begin{table}
\caption{Job Classes on Polaris}
\begin{center}
\begin{tabular}{lp{0.9in}lc}
\toprule
\textbf{Job Size} & \textbf{Queues} & \textbf{Nodes} & \textbf{Max Walltime} \\
\midrule
Small & \texttt{small, small-backfill} & 10--24 & 3 hrs \\
Medium & \texttt{medium, medium-backfill} & 25--99 & 6 hrs \\
Large & \texttt{large, large-backfill} & 100--496 & 24 hrs \\
\bottomrule
\end{tabular}  
\end{center}
\label{queue-table}
\end{table}

\subsection{System Data}
\begin{figure*}[htbp]
\centerline{\includegraphics[width=4.5in]{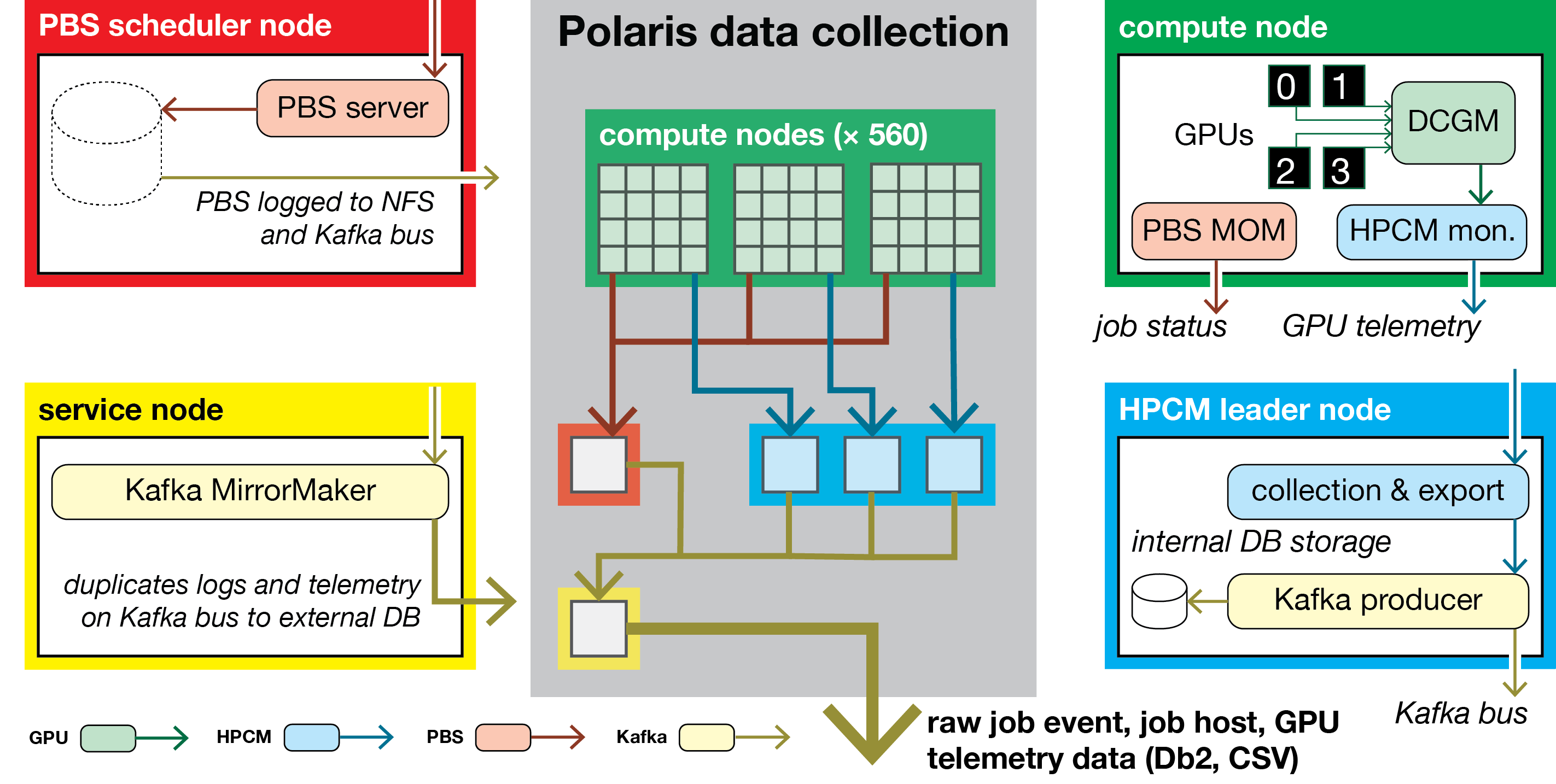}}
\caption{Telemetry and job data collection flow on the Polaris supercomputer. Compute node GPU metrics are sampled by NVIDIA and HPCM services, and job event data are recorded on the scheduler node. These data are streamed via a Kafka bus for later en masse processing.}
\label{data_collection}
\end{figure*}

Polaris continuously gathers information on itself using multiple layers of data communication and aggregation. 
This data is then transferred from its source to a designated, downloadable location.
Figure \ref{data_collection} depicts this process.

\begin{table*}
\caption{Data Summary}
\begin{center}
\begin{tabular}{lp{3in}rr}
\toprule
\textbf{Data} & \textbf{Contents} & \textbf{Annual Lines} & \textbf{Annual Size} \\
\midrule
\textbf{Original Data:} & & & \\
PBS Job Data & Job metadata, one row per job & 80 million & 7 GB \\ 
PBS Host Data & Mapping of hosts (nodes) to job, one row per host & 80 million & 7 GB \\ 
GPU Telemetry Data & Many measurements, one row per node per timeslice (5 sec) & 3.2 billion &  800 GB \\ 
\midrule
\textbf{Processed Data:} & & & \\
Complete Preprocessed Data & All original data in a single schema (no resolution loss) & 3.2 billion & 800 GB \\
Per-Job Summary Data & Compact, rich data for each job (analysis-ready) & 30 million & 48 GB \\
\bottomrule
\end{tabular}  
\end{center}
\label{data-table}
\end{table*}

In this work, we collect and examine three datesets.
\textbf{PBS Job Data} is collected by the PBS Pro scheduler node and includes detailed job characteristics such as the user, project, submission, and completion timestamps, the requested number of nodes, runtime, and assigned queue of each job.
\textbf{PBS Host Data} is also collected by the PBS Pro scheduler node and maps job IDs to the nodes assigned for their execution.
Finally, \textbf{GPU Telemetry Data} contains rich metrics detailing GPU core utilization, memory usage and allocation, power consumption, and temperature.

In this study, we consider three telemetry metrics: (1) GPU utilization (percentage of time the GPU cores were active), (2) GPU memory utilization (percentage of time GPU memory was being read/written), and (3) GPU memory allocation. 
These metrics are sampled every five seconds by NVIDIA’s Data Center GPU Manager (DCGM) tools and collected by HPCM’s native monitoring system.

PBS and GPU telemetry data are routed from their original source via a Kafka bus \cite{ApacheKafka} to permanent Db2 storage \cite{IBMDb2} external to Polaris. The datasets we analyze span \textbf{from January 1, 2024, to December 31, 2024} and are published in three schemas as outlined in Table \ref{data-table}. 
Tables  \ref{nvidia-terms} and \ref{derived-terms} list the sample and derived metrics used throughout the paper.

\begin{table*}
\caption{Taxonomy of Sampled Metrics}
\begin{center}
\begin{tabular}{lllp{3in}}
\toprule
\textbf{Conventional Term} & \textbf{HPCM Metric} & \textbf{NVML API Variable \cite{nvml}} & \textbf{Description \cite{nvml}} \\
\midrule
GPU utilization & \verb|GPU_load| & \verb|nvmlUtilization_t::gpu| & Percent of time over the past sample period during which one or more kernels were executing on the GPU \\
GPU memory utilization & \verb|GPU_mem_util| & \verb|nvmlUtilization_t::memory| & Percent of time over the past sample period during which global (device) memory was being read or written \\
GPU memory allocation & \verb|GPU_mem_alloc| & \verb|nvmlMemory_t::used| & Sum of reserved and allocated device memory (in bytes) \\
\bottomrule
\end{tabular}  
\end{center}
\label{nvidia-terms}
\end{table*}

\begin{table*}
\caption{Taxonomy of Derived Metrics}
\begin{center}
\begin{tabular}{lp{4in}l}
\toprule
\textbf{Derived Metric} & \textbf{Description}  & \textbf{Range}\\
\midrule
\verb|gpu_count| & Max number of GPUs with nonzero GPU utilization on any one node in a job & [0, 4] GPUs\\
\verb|num_nodes| & Number of nodes allocated to a job & [10, 496] nodes\\
\verb|node_hours| & Job node count $\times$ job runtime in hours & [0, 11,904] node-hours \\
\verb|runtime_seconds| & Job runtime in seconds & [0, 86,400]seconds\\
\midrule
\verb|total_load_mean| & Total node GPU utilization averaged across all nodes in a job & [0, 100] \%\\
\verb|total_mem_mean| & Total node GPU memory utilization averaged across all nodes in a job & [0, 100] \%\\
\verb|total_alloc_pct_mean| & Overall ratio of allocated GPU memory to total memory capacity, averaged across all nodes in a job & [0, 100] \%\\
\verb|total_power_mean| & Total node GPU power averaged across all nodes in a job & Watts (W)\\
\verb|total_energy| & Integral under instantaneous GPU power curve, summed for all nodes in a job & Kilojoules (KJ)\\
\midrule
\verb|ri_spatial_load| & $RI_{spatial}$ applied to GPU utilization & [0, 1]\\
\verb|ri_spatial_mem_util| & $RI_{spatial}$ applied to GPU memory utilization& [0, 1]\\
\verb|ri_spatial_mem_alloc| & $RI_{spatial}$ applied to GPU memory allocation & [0, 1]\\
\verb|ri_temporal_load| & $RI_{temporal}$ applied to GPU utilization & [0, 1]\\
\verb|ri_temporal_mem_util| & $RI_{temporal}$ applied to  GPU memory utilization & [0, 1]\\
\verb|ri_temporal_mem_alloc| & $RI_{temporal}$ applied to  GPU memory allocation & [0, 1]\\
\bottomrule
\end{tabular}  
\end{center}
\label{derived-terms}
\end{table*}

\section{Data Co-Analysis}\label{data-section}
Analyzing system-wide GPU telemetry data and scheduler logs presents unique challenges. The schemas vary significantly, making data integration complex and expensive, and the large volume of data further complicates efficient processing and analysis.

\begin{figure}[b] 
\centerline{\includegraphics[height=4in]{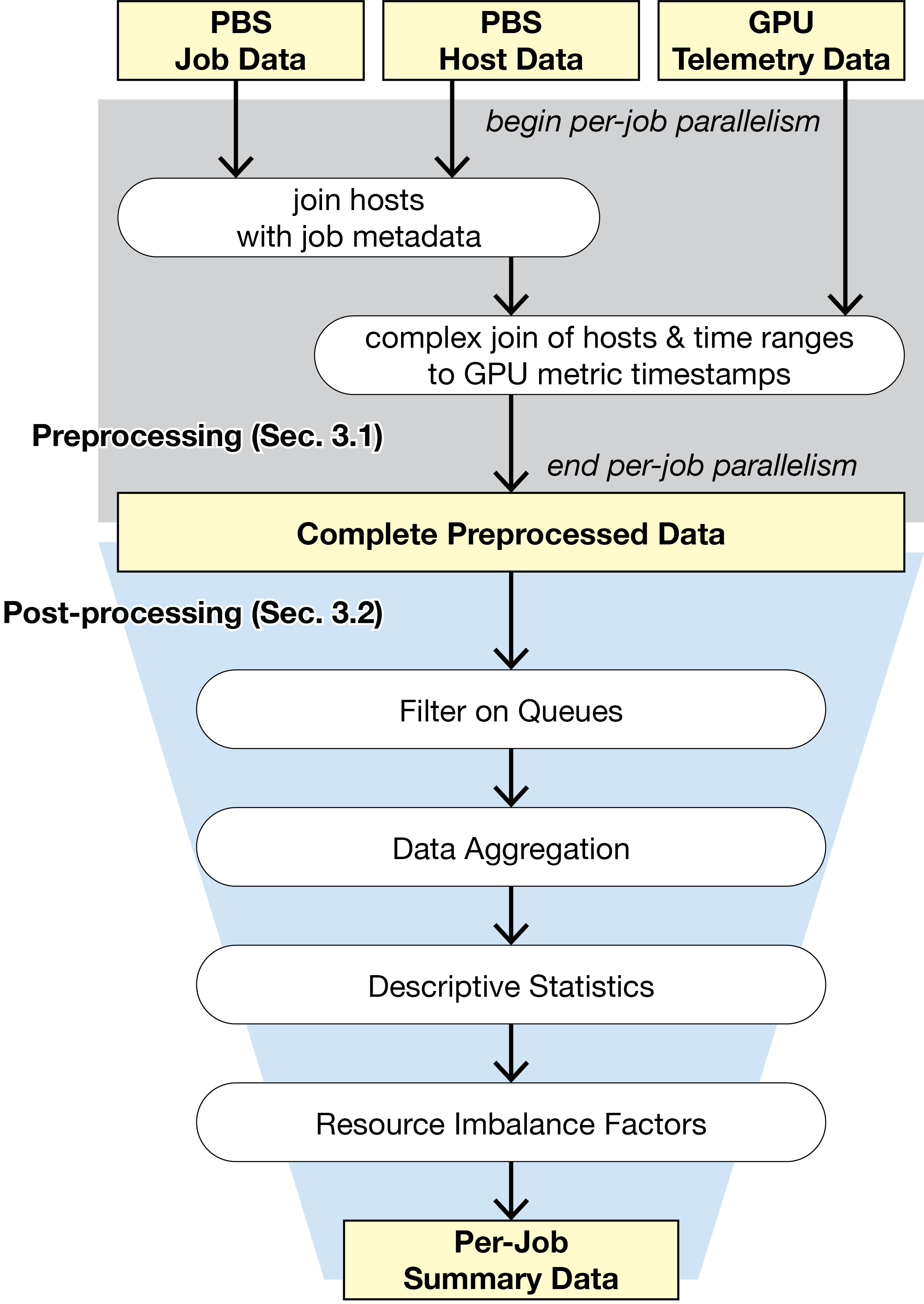}}
\caption{Our data co-analysis process:
(i) Preprocessing stage (gray) and
(ii) Post-processing stage (blue). 
}
\label{process-data-diagram}
\end{figure}

Our co-analysis workflow is presented in Figure \ref{process-data-diagram}. 
Through a series of pre- and post-processing steps, we create a unified summary dataset that enables system-wide pattern analysis. Pre-processing consolidates all original data into a common schema, while post-processing generates a condensed, analysis-ready dataset. 

The co-analysis workflow is implemented in Python using well-established tools: Pandas DataFrames \cite{mckinney2010data}, zstd compression \cite{collet2016zstd}, and Numba \cite{lam2015numba}. 
While more sophisticated data frameworks exist, using these standard tools allows us to make a codebase that is easily modified, optimized, and shared. 
The software tool (\textasciitilde500 lines of code), along with the associated datasets, will be released as open-source code on GitHub [link]. 

The remainder of this section describes our pre- and post-processing steps, detailing the challenges faced and strategies employed to overcome them.
These challenges are common when analyzing supercomputer system logs, regardless of the system architecture, data schema, or data collection tools employed.
\emph{Our goal is that our explanation, together with the release of the data and our open-source code, provides a clear and adaptable foundation for others to reproduce our analysis or tailor it to their use case}.

\subsection{Preprocessing}\label{preprocessing}

Preprocessing in this study joins the disperate datasets into a single schema.
To enable effective analysis, we correlate resource usage with job attributes by integrating data from PBS job logs and GPU telemetry--- a challenging task due to their size and differing schemas.
The top of Figure \ref{process-data-diagram} illustrates this procedure.

Table \ref{data-table} contains descriptions of the original datasets.
In this data, associating jobs with their assigned nodes is a straightforward join on job identifiers. However, linking job metadata to GPU telemetry is more complex. Each telemetry row contains various metrics (e.g., temperature and power), a timestamp, and the node (host) identifier---but it does not reference the specific job running on that node at that time. To resolve this, we determine which job was running on a given node by aligning timestamps from the telemetry data with time ranges in the PBS job data.

While trivial on a small scale, our study processes a full year of data---collected from 560 nodes---with a timestamp resolution of five seconds. This results in high computational demands. 
To address this challenge, we leverage parallelism  and checkpointing.
We use Numba and built-in Pandas mapping calls to facilitate parallelism, effectively assigning each worker a job identifier to process.
We also checkpoint progress in one-month chunks to avoid data loss, and we use compression to both save space and improve data load times.
The final output is the \textbf{\textit{complete preprocessed data}}, where each timestamp provides a comprehensive view of the system and job attributes.

\subsection{Post-processing}\label{statistics}
While the complete preprocessed data offers high-resolution details, the scale complicates system-wide analysis.
This data is post-processed to condense the contents while preserving critical insights and system-wide relevance.
In our approach, aggregations and statistical summaries are applied to enhance interpretability, again leveraging standard library calls and parallelism whenever possible.
The resulting dataset—of which a subset is used throughout this paper—is reduced by 94\% compared to the complete preprocessed data, significantly benefiting analysis efficiency.
The bottom of Figure \ref{process-data-diagram} illustrates this process. 

\textbf{Data Aggregation.} Several values are derived from the raw data. The number of GPUs a job uses, ranging from 0 to 4, is determined by counting the maximum number of GPUs on any node with a nontrivial ($>2\%$) maximum GPU utilization. GPU energy is calculated as the integral with respect to time under the instantaneous, node-wide GPU power curve for each node assigned to a job. Node-hours are computed as the number of nodes in a job multiplied by its total runtime (in hours). 

\textbf{Descriptive Statistics.}
We summarize key characteristics of the preprocessed data, including minimum, maximum, mean, and standard deviation. Pearson correlation coefficients between these values are also calculated to facilitate correlation analysis \cite{Pearson1895}. We analyze the dataset at two levels: \emph{node level} (e.g., mean power consumption for node X in job Y) and \emph{job level} (e.g., mean power consumption across all nodes in job Y). This dual approach captures both node-specific trends and overall job behavior.

\textbf{Resource Imbalance.}
A resource imbalance ($RI$) coefficient quantifies disparities in resource utilization within a system, measuring consistency in inter- or intra-node resource usage. Initially introduced by Peng et al. to analyze memory usage patterns \cite{Peng2022} and later extended by Li et al. to other metrics such as CPU load \cite{Li2023}, we adopt this method to assess resource utilization consistency.


For a job running on $N$ nodes for a duration $T$,  $RI$ is derived from a telemetry measurement $U$ sampled from node $n$ at a time $t$. 
Higher $RI_{temporal}$ suggests more erratic intra-node behavior, while higher $RI_{spatial}$ indicates greater variability across nodes.

\begin{itemize}
    \item \textbf{Temporal Imbalance} ($RI_{temporal}$): Measures resource variability within a node during a job's runtime. It ranges from $[0, 1]$, where 0 indicates constant usage and 1 indicates high fluctuation. $RI_{temporal}$ for a given job is the maximum $RI_{temporal}$ value on any of its assigned nodes, calculated using:
    \begin{equation}\label{ri-temporal} RI_{temporal}(r) = \max_{1 \le n \le N} \left(1 - \frac{\sum_{t=0} ^{T} U_{n,t}} {\sum_{t=0} ^{T} \max_{0 \le t \le T} U_{n,t}}\right) \end{equation}

    \item \textbf{Spatial Imbalance} ($RI_{spatial}$): Measures consistency across nodes within a job. Also ranging from $[0, 1]$, a value of 0 indicates identical resource usage across all nodes, while 1 represents entirely independent behavior, calculated using:
    \begin{equation}\label{ri-spatial} RI_{spatial}(r)=1-\frac{\sum_{n=1} ^N \max_{0 \le t \le T} (U_{n,t})}{\sum_{n=1} ^ N \max_{0 \le t \le T,1 \le n \le N} (U_{n,t})}
\end{equation}   
\end{itemize}





\begin{figure}
    \centering
    \includegraphics[width=3in]{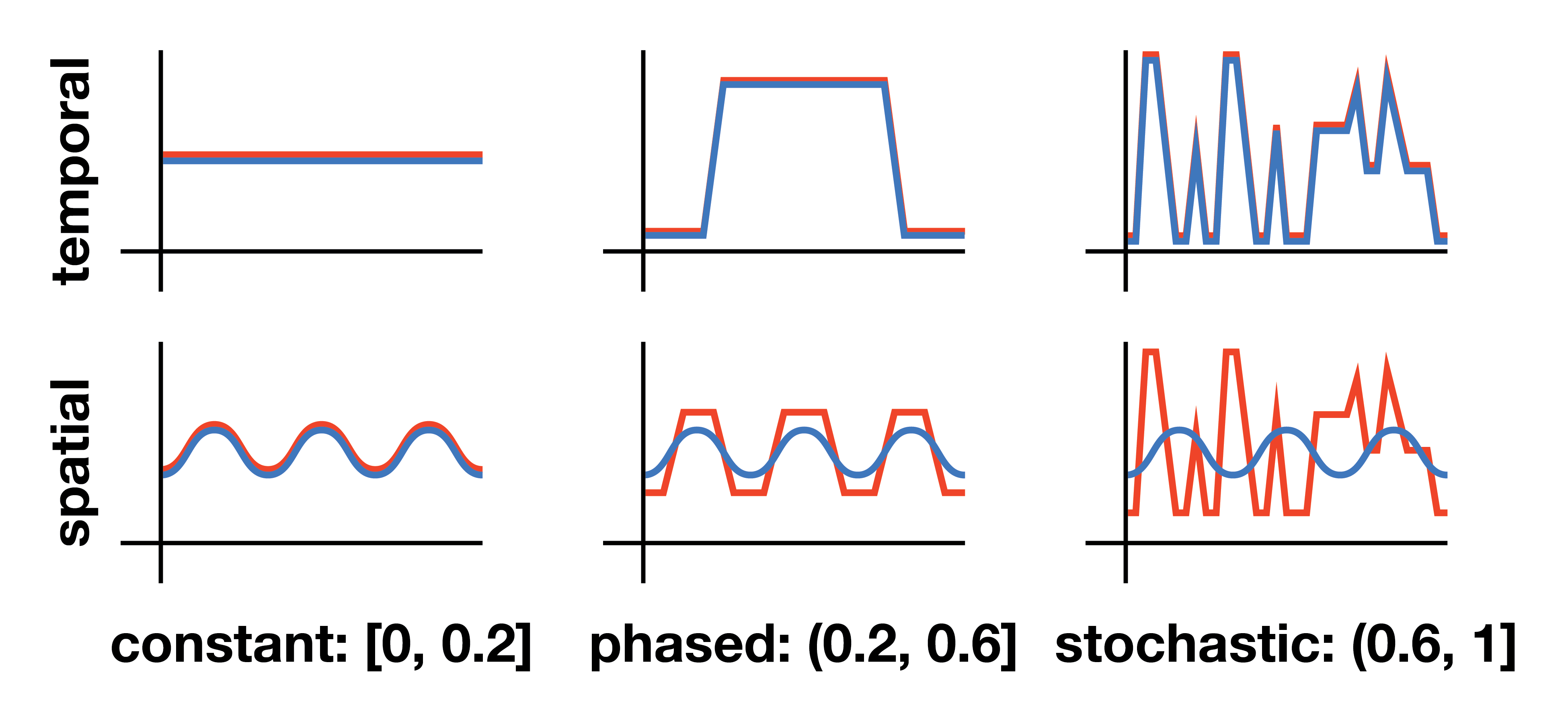}
    \caption{Resource imbalance categories showing  resource demands on two nodes (blue and red).
    (Top) $RI_{temporal}$ captures intra-node variation---constant workloads exhibit no change, while stochastic workloads show significant fluctuations. 
    (Bottom) $RI_{spatial}$ captures inter-node differences---constant workloads have uniform resource usage, while stochastic workloads display high node-to-node variance.
    }
    \label{toy-examples}
\end{figure}

We classify $RI$ values into three range categories: \emph{constant} [0, 0.2], \emph{phased} (0.2, 0.6], and \emph{stochastic} (0.6, 1], illustrated in Figure \ref{toy-examples}. 
$RI_{temporal}$ coefficients concern the resource's variation over time for the \textit{most} inconsistent node. Constant jobs show little change, phased jobs have stable periods interspersed with fluctuations, and stochastic jobs exhibit significant noise.
$RI_{spatial}$ coefficients concern the resource’s variation across different nodes within a job. Spatially constant jobs demonstrate uniform usage across nodes, phased jobs vary in amplitude or pattern, and stochastic jobs exhibit little correlation among nodes.

This study considers three resources: \textbf{GPU utilization, GPU memory utilization, and GPU memory allocation}. We compute six $RI$ coefficients per job, $RI_{temporal}$ and $RI_{spatial}$ for each of the three resources.

\begin{figure}
    \centerline{\includegraphics[width = 3.5in]{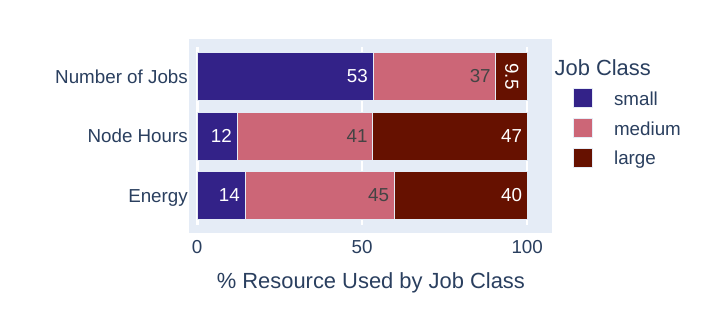}}
    \hfill
    \centerline{\includegraphics[width = 3.5in]{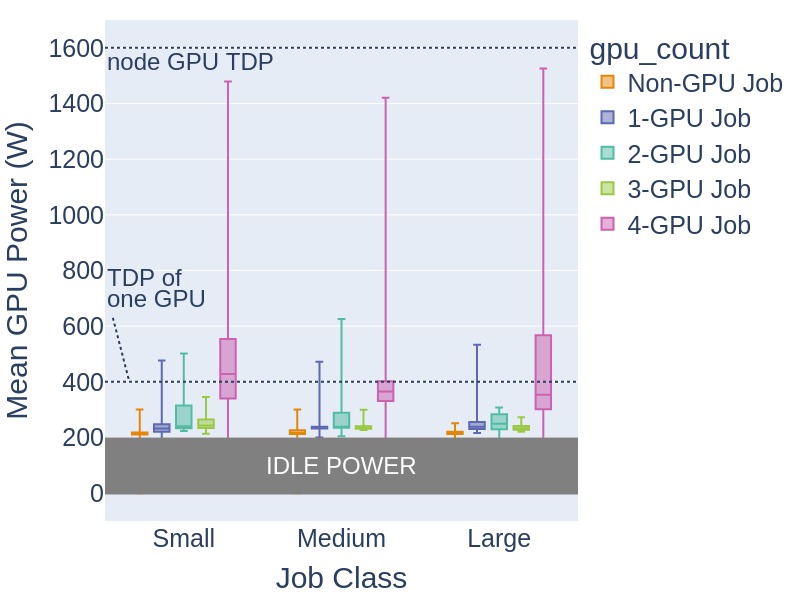}}
    \caption{ 
    (Top) Resource decomposition by job class. 
    (Bottom) Mean power draw by job class (see Table \ref{queue-table}) and number of GPUs used, showing the median, quartiles (Q1 and Q3) as box edges, minimum and maximum values as whisker endpoints, and outliers along an outer line. 
    }

    \label{resource-breakdown}
\end{figure}


\section{Job Characteristics}\label{job-section}

Polaris operates on a batch-scheduled, bare-metal model.
Since a job reserves the entire node, characterizing jobs by their attributes is essential for understanding system demands. We distinguish jobs by class, defined by the size (binned according to the node count, see Table \ref{queue-table}), and by the number of GPUs utilized. Key job attributes---including runtime, node hours, node-wide GPU energy, and node-wide GPU power---are analyzed.

The top of Figure \ref{resource-breakdown} shows system demand by job class. Most Polaris jobs are small (10--24 nodes), and predictably medium and large jobs dominate node-hours and GPU energy. 

The bottom of Figure \ref{resource-breakdown} further examines node GPU power by job class. 
Each Polaris node contains four A100 GPUs, with an idle power draw of \textasciitilde50 W per GPU and a maximum TDP of \textasciitilde400 W.
This results in a total node-wide power range of about 200 W (idle) to 1,600 W (max TPD). 
Figure \ref{resource-breakdown} illustrates power consumption by job class and the number of GPUs used. 
From this, we derive two key insights:
First, GPU power usage is surprisingly low, averaging just 400 W in the worst case---equivalent to the TDP of a single GPU, or just 25\% of a node’s maximum TDP.
Second, idle power is equivalent to \textit{half} of this 400 W draw. 

\begin{mybox}
\textbf{Observation 1.} GPU power is overprovisioned and inefficiently utilized: jobs typically consume just 400 W, roughly equal to the TDP of a single GPU. Additionally, at least 200 W of this is idle power, meaning half of the total draw comes from simply keeping the GPUs powered on, rather than from active usage.
\end{mybox}

\begin{figure}[htbp]
\centerline{\includegraphics[width = 3.2in]{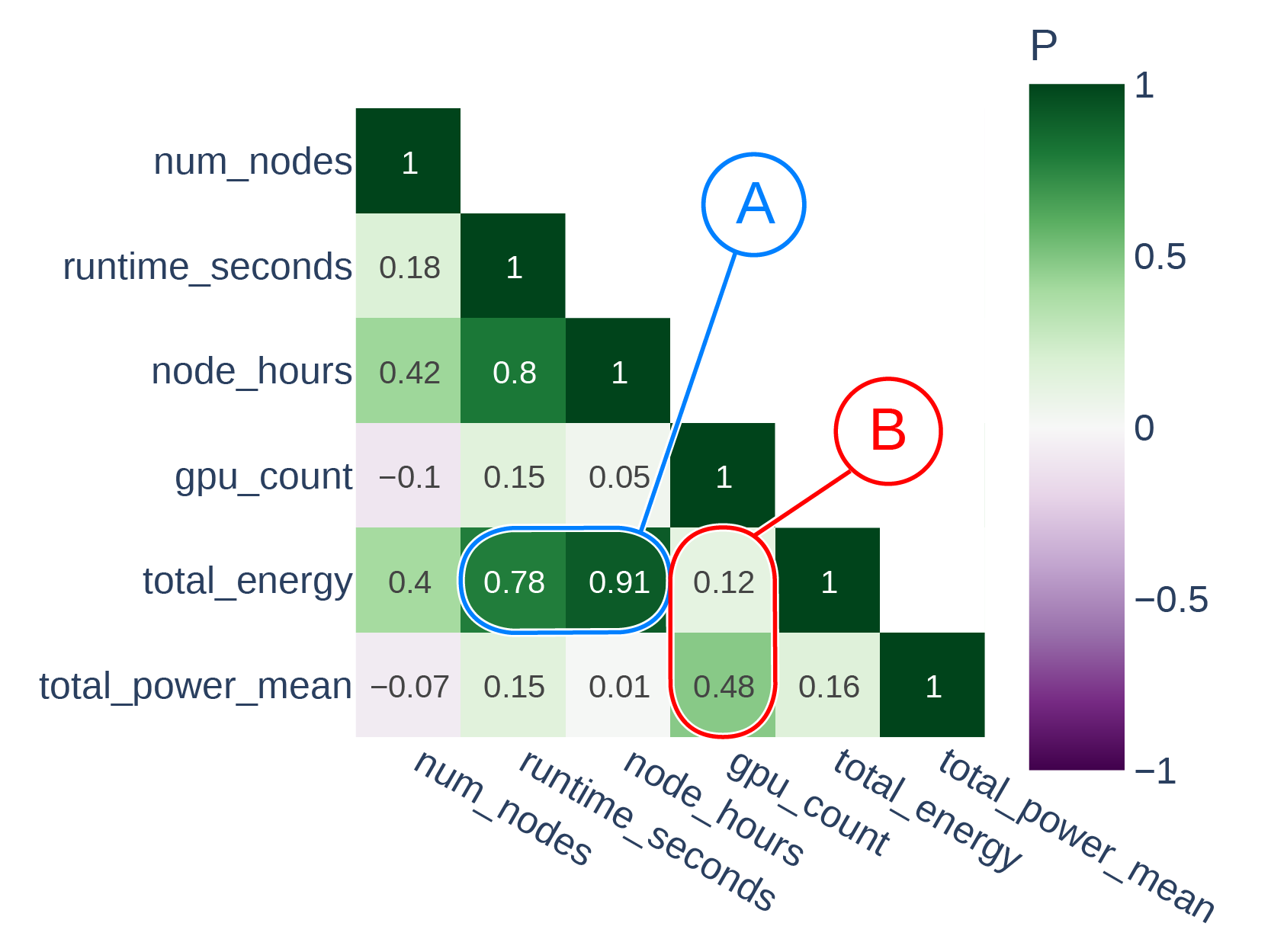}}
\caption{Pearson’s coefficients for job metrics.  
Node hours, runtime, and energy show strong correlations (A), while GPU count, power, and energy correlate weakly (B).}
\label{pearson-system}
\end{figure}

Figure \ref{pearson-system} shows a heatmap of Pearson correlation coefficients for various job metrics, revealing more key trends. First, a \textit{strong} correlation exists among energy, node hours, and runtime (A). And second, only a \textit{weak} correlation is observed between GPU power and GPU count (B).

\begin{mybox}
\textbf{Observation 2.} Jobs that use more GPU power do not necessarily use more GPUs, as the impact of idle power significantly outweighs the power draw from active usage.
\end{mybox}

\textbf{Discussion.} Polaris nodes are power-oversubscribed, consuming only 25\% (400 W) of their maximum GPU TDP (1600 W) on average. Half of this draw is idle power (200 W), and this pattern is consistent across all job classes and GPU counts. While numerous GPU power-saving strategies exist, our analysis indicates that prioritizing idle power reduction could lead to substantial savings, representing a clear opportunity for system-wide optimization.
Potential approaches to test and consider, for example, could include enabling software GPU power states (P-states, \cite{pstate}), advanced power capping \cite{10820672}, or co-location of GPU applications \cite{10.1145/3547276.3548630}.

\begin{oppbox}
\textbf{Opportunity 1.} 
Current GPU power provisioning leads to underutilization, with jobs consuming only 25\% of the total available GPU power on average.
While there are many approaches to reducing GPU power, our analysis suggests that focusing on idle power reduction could yield significant benefits.
\end{oppbox}

One potential approach to power optimization deserves some discussion.
NVIDIA's NVAPI defines \textbf{P-states} (performance states), ranging from P0 to P15, as settings to control power optimization based on workload demands \cite{nvidiaNVAPI}.
Consumer-grade GPUs, such as NVIDIA's Optimus series \cite{nvidia_optimus_policies}, leverage P-states as a standard feature for power savings, reducing idle power in scenarios like DVD playback or an idle display \cite{nvidia_nvapi_pstate}.
Advanced architectures leverage P-states too, including the Tesla T4 series, which offers P0 and P8 states. For example, this has been used to enhance energy-efficiency tuning for AI workloads \cite{10818209}.
However, NVIDIA's A100, a workhorse of enterprise systems like Polaris, lacks this software-controlled power management---a limitation for power optimization. 
A100s are locked at P-state P0 (``Maximum Performance''\cite{nvml}), preventing all scales of automated power reductions---and particularly preventing power conservation intended to lower idle power costs.

The absence of P-state options for A100 GPUs significantly limits  opportunities to improve power efficiency.
If NVIDIA were to introduce granular power management across all architectures, enterprise systems (such as data centers and HPC systems) could better-balance performance and energy conservation. 

\begin{oppbox}
\textbf{Opportunity 2.} 
While many consumer and some enterprise NVIDIA GPUs utilize software-controlled P-states for power efficiency, A100s currently lack this capability. Enabling P-state control---either now or in future GPU designs---offers a promising avenue for reducing idle power consumption and operational costs.
\end{oppbox}

\section{GPU Utilization}\label{gpu-load-section}
Each Polaris node is equipped with four GPUs, and jobs have exclusive access to all four devices on every assigned node throughout their runtime. 
Thus, we characterize jobs based on how intensely they utilize their allocated  GPUs. We define GPU utilization (see \verb|GPU_load| metric in Table \ref{nvidia-terms}) as the \textit{proportion} of time during which any kernel executes on the GPU over a given sample period.

\begin{figure}[htbp]
\centerline{\includegraphics[width = 3.1in]{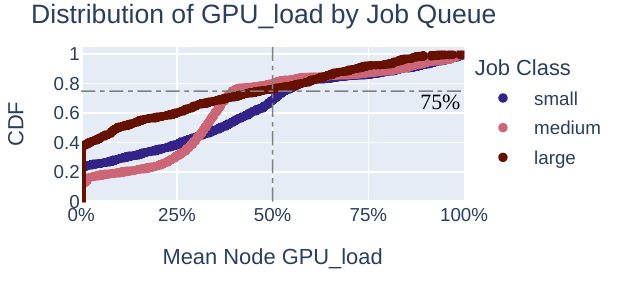}}
\caption{Distribution of mean GPU utilization.
}
\label{cdf_load}
\end{figure}

\subsection{GPU Utilization by Job Class}
Node GPU utilization is defined such that 100\% utilization indicates that all GPUs on all nodes were constantly executing kernels throughout the job's runtime.
Figure \ref{cdf_load} shows the cumulative distribution function (CDF) \cite{casella2002statistical} of GPU utilization, with each job class represented by a different color. The plot clearly identifies GPU underutilization in the average case.

\begin{mybox}
\textbf{Observation 3.} On average, approximately 75\% of all Polaris jobs utilize GPUs for half or less of their runtime.
\end{mybox}

\subsection{$RI$ Coefficients for GPU Utilization}\label{load-ri-subsection}

\begin{figure}
    \centerline{\includegraphics[width = 2.8in]{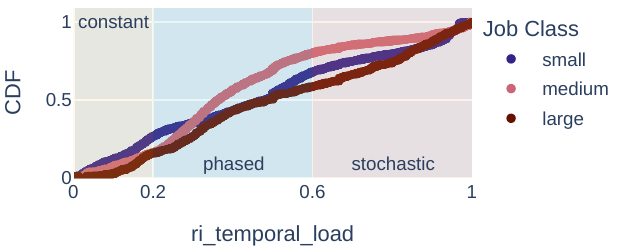}}
    \centerline{\includegraphics[width = 2.8in]{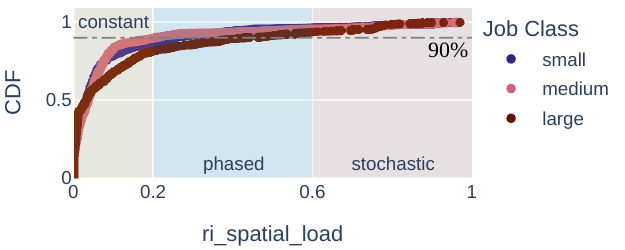}}
    \caption{(Top) $RI_{temporal}$ for mean GPU utilization. 
    (Bottom) $RI_{spatial}$ for mean GPU utilization. 
    }
    \label{ri_load}
\end{figure}

\begin{figure}[htbp]
\centerline{\includegraphics[width = 3.2in]{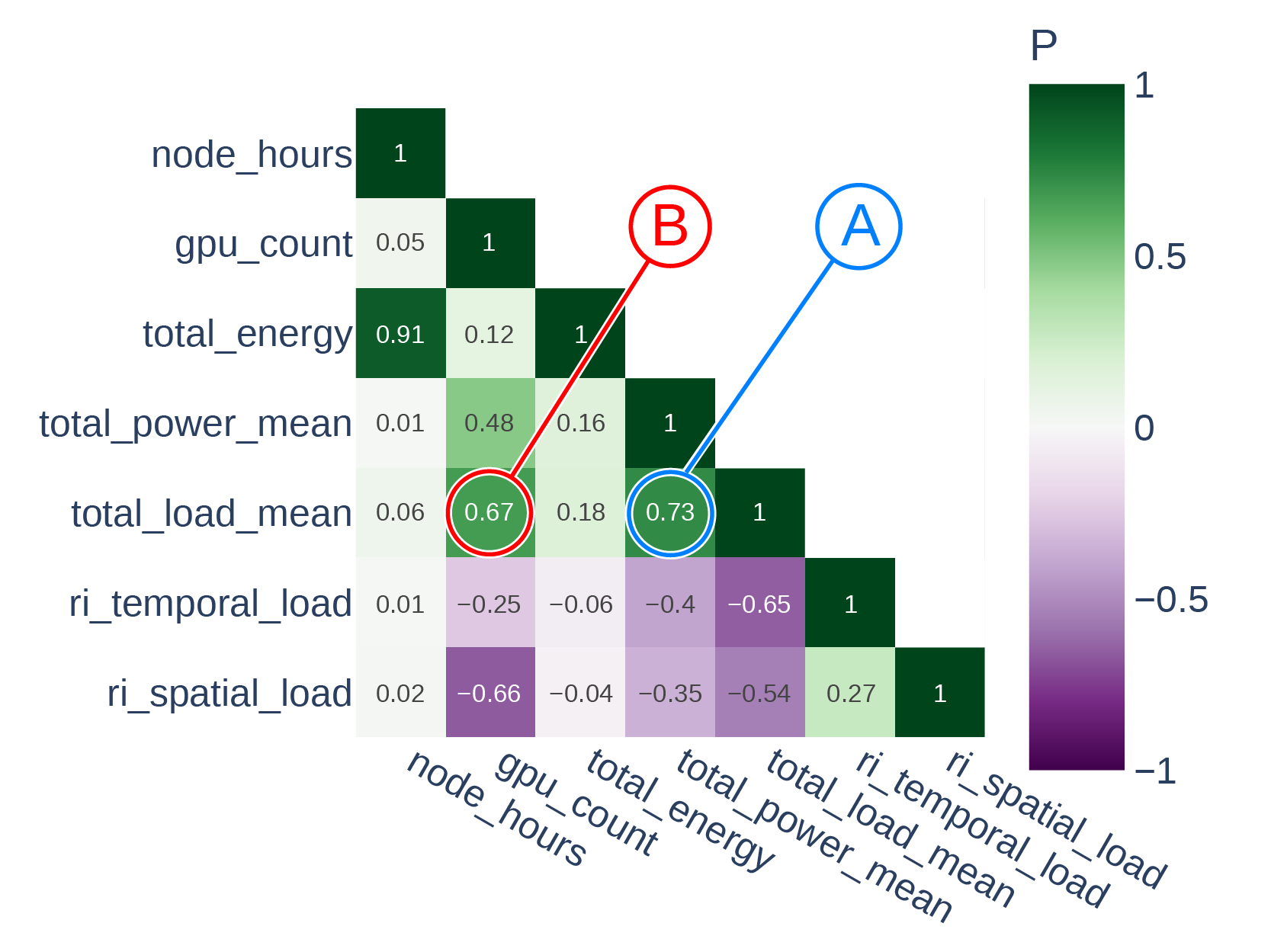}}
\caption{Heatmap of GPU utilization, job statistics, and RI coefficients. 
Jobs that show higher mean GPU utilization typically consume more power and use more GPUs (A) and tend to use more GPUs (B).
}
\label{heat_load}
\end{figure}

To characterize GPU utilization, we compute $RI_{temporal}$ to measure usage consistency on individual nodes, and $RI_{spatial}$ to assess utilization similarity across nodes within a job. 

Figure \ref{ri_load} shows the CDF for $RI_{temporal}$ and $RI_{spatial}$ applied to GPU utilization.
Within a single node, jobs of all classes exhibit varying $RI_{temporal}$ values, meaning GPU usage fluctuates. In fact, jobs are about equally likely to fall into any of the three categories (constant, phased, stochastic). However, across nodes, jobs of any class tend to have low $RI_{spatial}$ values, with more than 90\% of jobs falling into the constant category.

\begin{mybox}
\textbf{Observation 4.} Jobs exhibit variation in GPU utilization within a node, but they generally use GPUs consistently across all nodes within a job.
\end{mybox}

\subsection{Correlation of GPU Utilization, $RI$ Coefficients, and Job Attributes}\label{load-correlation-subsection}

Figure \ref{heat_load} presents a heatmap of Pearson correlation coefficients between key metrics.
Here, we reinforce some intuitive relationships: jobs with higher mean GPU utilization tend to consume more power (A), and higher GPU utilization also correlates to more GPUs being used (B).


\subsection{Discussion}
Our analysis of GPU utilization reveals substantial underutilization: approximately 75\% of jobs use GPUs for half or less of their runtime. Within a node, utilization often varies over time, with many jobs exhibiting irregular or bursty behavior. However, comparing nodes within the same job, GPU usage tends to be consistent.

This consistency highlights a key opportunity: job-level power management. Because nodes within a job behave similarly, power saving strategies can be applied uniformly at the job level, and sampling can be performed on one or a small subset of nodes. This makes job-level power tuning a practical and scalable way to better match power allocation to actual workload demands.

\begin{oppbox}
\textbf{Opportunity 3.} Most jobs underutilize GPUs, presenting a clear opportunity for power optimization. However, selecting and applying the right power management strategy can be complex and costly. Since GPU usage is consistent across nodes within a job, strategies can be applied at the job level, with any runtime sampling requirements taken from only a single node. This offers scalability and ease of implementation.
\end{oppbox}

\section{GPU Memory Utilization}\label{mem-util-section}
Similar to GPU utilization, jobs can be categorized based on the intensity of their GPU memory utilization. This metric (\verb|GPU_mem_util| in Table \ref{nvidia-terms}) measures the \emph{proportion} of time within a sampling period during which some read or write operation is active.

\subsection{GPU Memory Utilization by Job Class}  
\begin{figure}[htbp]
\centerline{\includegraphics[width = 3.1in]{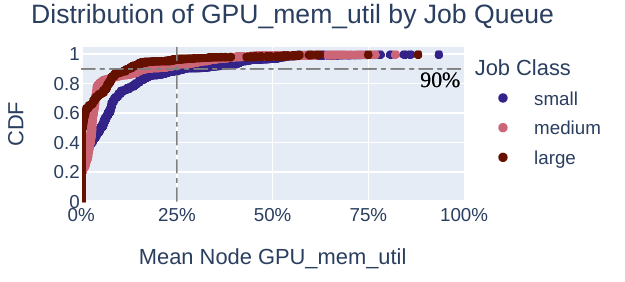}}
\caption{Distribution of mean GPU memory utilization.
}
\label{cdf_mem}
\end{figure}

Total GPU memory utilization is defined such that 100\% indicates GPUs on all nodes within a job were actively performing read and/or write instructions during the job's runtime.
Figure \ref{cdf_mem} presents the CDF of mean GPU memory utilization, with each job class represented by a different color. The plot reveals minimal memory usage across all job types.

\begin{mybox}
\textbf{Observation 5.} GPU memory is underutilized: on average, nearly 90\% of all Polaris jobs access GPU memory for less than a quarter of the job's runtime.
\end{mybox}

\subsection{$RI$ Coefficients for GPU Memory Utilization} \label{ri-mem-section}  

\begin{figure}
    \centerline{\includegraphics[width = 2.8in]{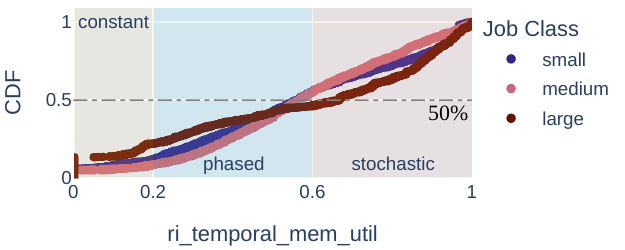}}
    \centerline{\includegraphics[width = 2.8in]{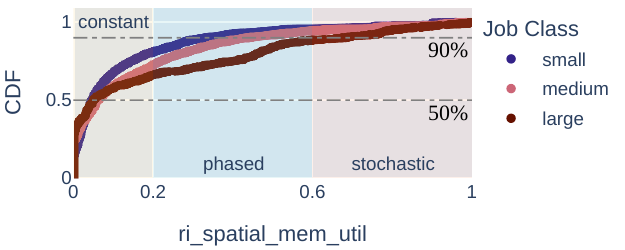}}\
    \caption{(Top) $RI_{temporal}$ for mean memory utilization. 
    (Bottom) $RI_{spatial}$ for mean memory utilization. 
    }
    \label{ri_mem}
\end{figure}

For GPU memory utilization, $RI_{temporal}$ coefficients measure the consistency of read/write operations within a node, while $RI_{spatial}$ coefficients assess whether memory access patterns are similar across nodes within a job. 

Figure \ref{ri_mem} shows the CDF of $RI_{temporal}$ and $RI_{spatial}$ applied to GPU memory utilization.
Within a node, most jobs access memory inconsistently throughout runtime, with more than half of jobs falling into the stochastic temporal category.
Across nodes, jobs generally exhibit similar memory usage behaviors.
Very few jobs are spatially stochastic; more than half of all jobs are spatially consistent, and over 90\% fall into either the constant or phased categories.


\begin{mybox}
\textbf{Observation 6.} Jobs vary in GPU memory utilization within nodes, but they tend to read/write memory similarly across all nodes in the job.
\end{mybox}

\subsection{Correlation of GPU Memory Utilization, $RI$ Coefficients, and Job Attributes}  
\begin{figure}[t] 
\centerline{\includegraphics[width = 3.2in]{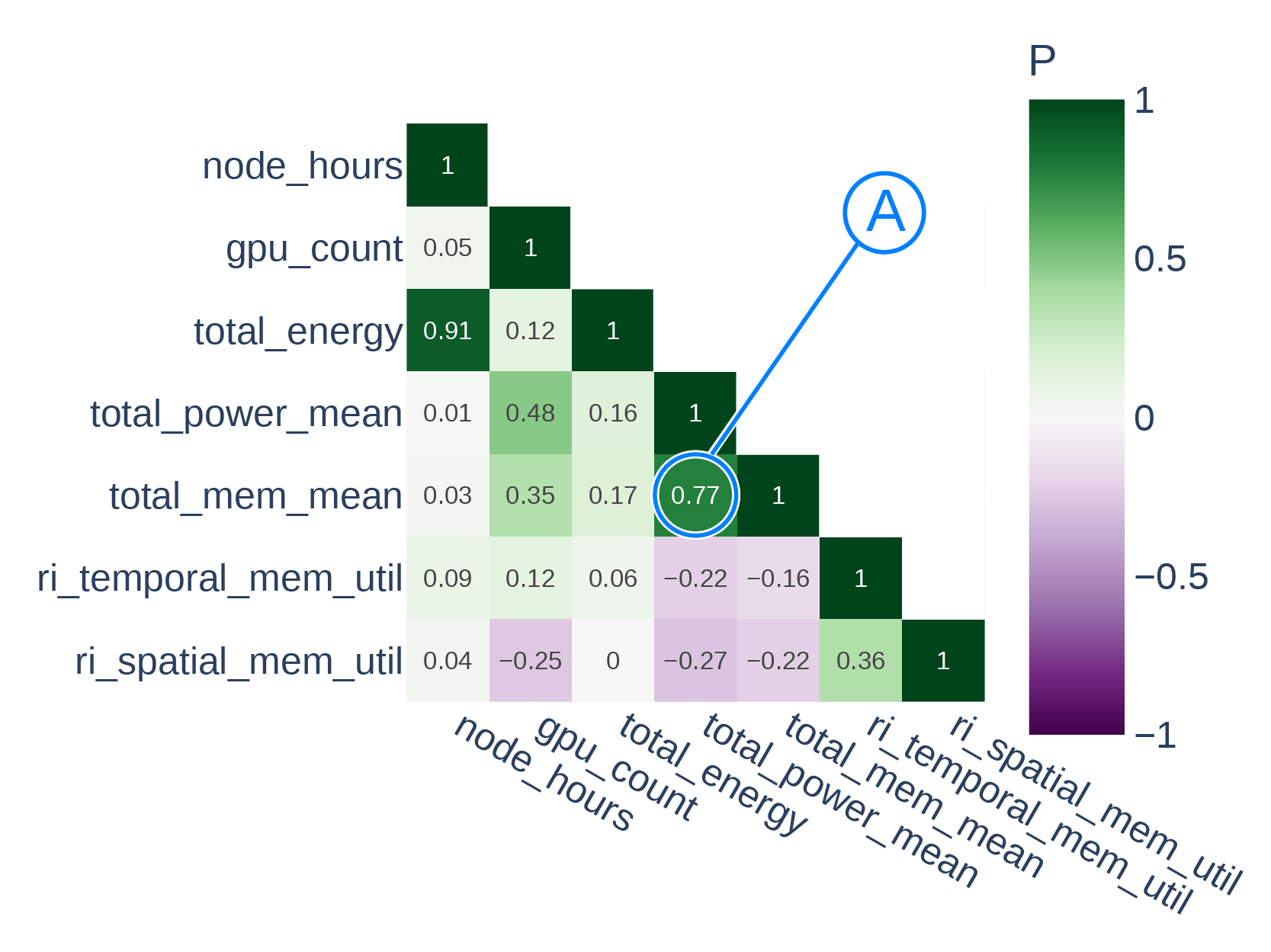}}
\caption{Heatmap of GPU memory utilization and RI coefficients. 
GPU memory utilization correlates with increased power demands (A).
}
\label{heat_mem}
\end{figure}

Integrating insights from GPU memory utilization trends, job attributes, and the corresponding $RI$ coefficients is challenging. To provide a concise view, Figure \ref{heat_mem} provides a heatmap of Pearson correlation coefficients among key metrics.
From this, we observe a clear trend indicating jobs with higher GPU memory utilization tend to consume more instantaneous power. Prior studies on GPU power consumption have shown that memory-intensive kernels consume 20--30\% more energy than compute-intensive kernels \cite{7830508}, aligning with our findings.

\begin{mybox}
\textbf{Observation 7.} Jobs with more frequent GPU memory read-write operations typically require higher instantaneous power. This is consistent with the established high cost of memory utilization.
\end{mybox}

\subsection{Discussion}
Our analysis shows that GPU memory is accessed infrequently during execution: nearly 90\% of jobs read or write memory less than a quarter of the time. Within a node, access patterns are often bursty, with over half of jobs exhibiting temporally stochastic memory usage. However, across nodes in a job, memory behavior tends to be uniform, with more than 50\% of jobs categorized as spatially constant.

These patterns suggest that memory remains a valuable target for power optimization. As with GPU utilization, the consistency of resource use across the nodes within a job supports the use of job-level strategies. However, unlike GPU utilization, memory access is extremely infrequent. Any selected power savings strategy must either capitalize on bursty and inconsistent usage or be resilient to it without compromising performance.
These might include, for example, enabling dynamic memory clock gating \cite{4685421}, using smaller-capacity GPUs for memory-light workloads \cite{10.1145/3632775.3662830}, or again leveraging P-state settings \cite{pstate}.

\begin{oppbox}
\textbf{Opportunity 4.}
GPU memory is consistently underutilized across jobs, presenting a promising opportunity for power savings. With memory utilization patterns similar across nodes, job-level strategies are once again feasible --- especially ones which take advantage of bursty and infrequent memory access.
\end{oppbox}

\section{GPU Memory Allocation}\label{allocation-section}
\vspace{-1.5mm}
To cohesively understand memory usage patterns, we also analyze GPU memory allocation. Each Polaris node contains four A100 GPUs, providing a total of 160 GB of GPU memory.
Jobs can be categorized by their GPU memory allocation behaviors (see \verb|GPU_mem_alloc| from Table \ref{nvidia-terms}).
To maintain consistency with the previous metrics, we will express this as a percentage.

\begin{figure}[htbp]
\centerline{\includegraphics[width = 2.8in]{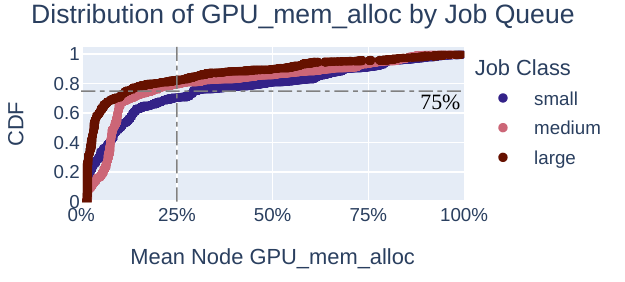}}
\centerline{\includegraphics[width = 2.8in]{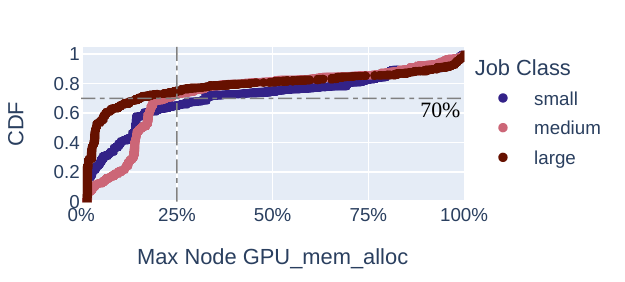}}
\caption{ Distribution of mean (top) and maximum (bottom) GPU memory allocation.
}
\label{cdf_alloc}
\end{figure}

\subsection{GPU Memory Allocation by Job Class}  
\vspace{-1.5mm}
GPU memory allocation is defined such that 100\% indicates all 160 GB of GPU memory across all nodes within a job were fully allocated for the duration of the job's runtime.
Since memory allocation is also a capacity concern, it is useful to consider both the average GPU memory allocation across all nodes in a job in addition to the maximum allocation from any one node in the job.
Figure \ref{cdf_alloc} presents the CDF of the GPU memory allocation per-job (both mean and max), with each job class represented by a different color.
The plot demonstrates that most jobs allocate relatively little GPU memory, with approximately 70\% allocating no more than 25\% of a node’s total capacity---both on average and at peak usage.

\begin{mybox}
\textbf{Observation 8.} GPU memory is underutilized: \textasciitilde70\% of all jobs allocate no more than a single GPU's memory capacity (40 GBs).
\end{mybox}

\subsection{$RI$ Coefficients for GPU memory allocation}
\vspace{-1.5mm}

\begin{figure}
    \centerline{\includegraphics[width = 2.8in]{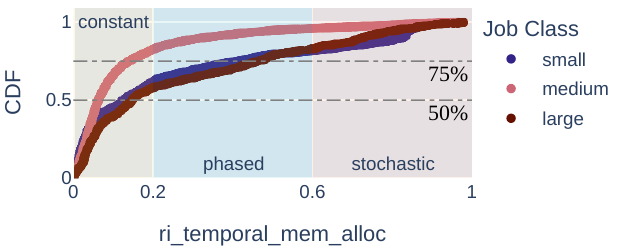}}
    \centerline{\includegraphics[width = 2.8in]{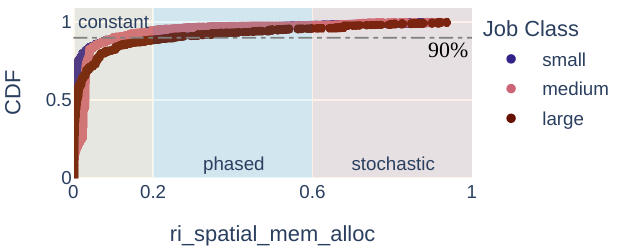}}
    \caption{(Top) $RI_{temporal}$ for memory allocation percent. 
  (Bottom) $RI_{spatial}$ for memory allocation percent. 
 }
    \label{ri_alloc}
\end{figure}

For GPU memory allocation, $RI_{temporal}$ coefficients quantify the consistency of memory allocation within a node, while $RI_{spatial}$ coefficients assess whether allocation patterns are similar across nodes within a job. 

Figure \ref{ri_alloc} presents the CDF of $RI_{temporal}$ and $RI_{spatial}$ for GPU memory allocation.
Over half of all jobs exhibit low temporal imbalance, placing them in the constant category.
Medium jobs are even more likely to be consistent, with over 75\%  categorized as constant.
All job classes spatially allocate memory consistently; more than 90\% of all jobs fall into the spatially constant category.

\begin{mybox}
\textbf{Observation 9.} GPU memory allocation and deallocation patterns are generally consistent within nodes and, more prominently, across all nodes in a job.
\end{mybox}

\subsection{Correlation of GPU memory allocation, $RI$ Coefficients, and Job Attributes}
\begin{figure}[t] %
\vspace{-0.3cm}
\centerline{\includegraphics[width = 3.5in]
{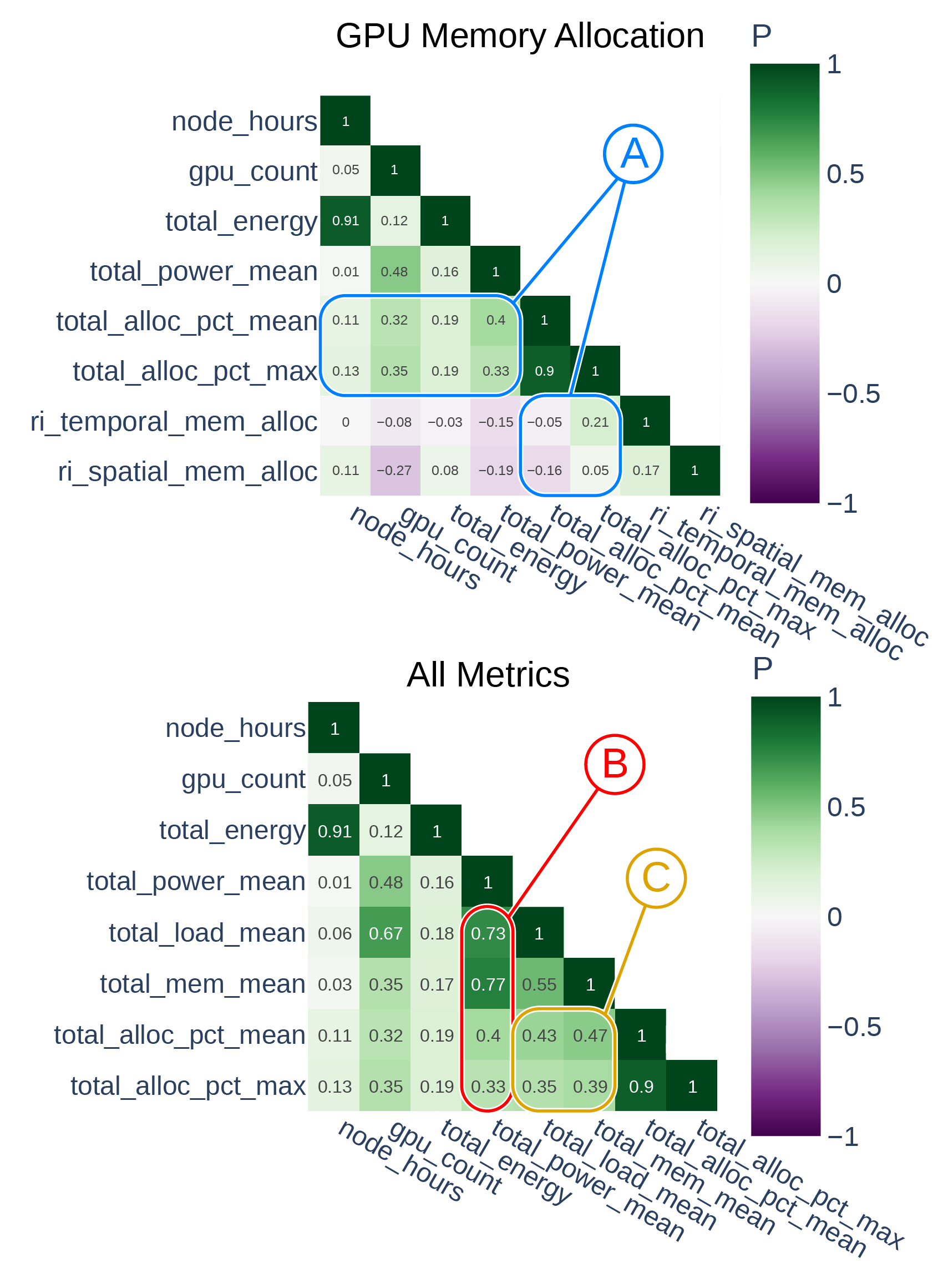}}
\caption{Heatmaps of GPU memory allocation and RI coefficients. (Top) Correlations between mean/max memory allocation, $RI$ coefficients, and job attributes show that memory allocation has little relationship with other metrics (A). (Bottom) Correlations between power and utilization metrics reveal that GPU and memory utilization contribute more significantly to power consumption than memory allocation (B), and further highlight the weak correlation between memory allocation and utilization (C).
}
\label{heat_alloc}
\end{figure}

\vspace{-1.5mm}
Integrating insights from GPU utilization trends, job attributes, and the corresponding $RI$ coefficients is challenging. To provide a concise view, Figure \ref{heat_alloc} provides heatmaps of Pearson correlation coefficients among key metrics. 
Looking at the top heatmap, we observe GPU memory allocation shows almost no correlation with other system and job attributes (A).
The lower heatmap highlights the comparatively minor impact of GPU memory allocation on power consumption, especially when contrasted with GPU utilization and GPU memory utilization (B).
We also observe very little correlation between GPU memory allocation and the other utilization metrics (C).
Notably, allocating more GPU memory does not strongly suggest that memory will be more highly-utilized.



\begin{mybox}
\textbf{Observation 10.}
GPU memory allocation has little correlation with other key metrics. Memory allocation is important for understanding capacity needs but offers limited insight into power efficiency.
\end{mybox}

\subsection{Discussion}
Our analysis shows that GPU memory capacity is frequently underutilized, with approximately 70\% of jobs allocating at most 40 GB (the memory capacity of a single GPU). Memory allocation patterns are generally consistent both over time and across nodes within a job. And while GPU memory allocation is important for capacity planning, it has minimal correlation with power consumption or utilization metrics. This decoupling simplifies system analysis by allowing power-focused studies to deprioritize allocation patterns. Optimization strategies centered on active memory usage are likely to yield more effective power reductions.

\begin{oppbox}
\textbf{Opportunity 5.}
GPU memory allocation does not directly correlate with GPU power usage. Optimizations that are utilization-aware are key approaches to reducing power consumption.
\end{oppbox}

\vspace{-2.5mm}
\section{Related Work}
\vspace{-1.5mm}
Hybrid CPU-GPU systems are now widely deployed in HPC due to their ability to handle a wide range of workloads. 
While log processing presents challenges, it remains a crucial tool for understanding and optimizing HPC system performance and efficiency. 

Various studies have analyzed log data for insights and system improvements. 
Shvets et al. \cite{shvets} examine how to effectively use supercomputer logs by integrating multiple profiles and data sources, which helps identify useful performance trends. 
Similarly, research on the MIT SuperCloud \cite{DT-hpca22} takes a broad view of workload characterization, highlighting low utilization rates and runtime variation as key areas of interest. 
Using data from NERSC, Li et al. \cite{Li2023} examine both system logs and workload benchmarking, incorporating the RI statistical definitions to provide a detailed view of system performance.
Long-term studies, such as the seven-year analysis of the Tian-Hai supercomputer \cite{iccs2018}, offer valuable insights into performance trends over extended periods. 
Additionally, a long-term log analysis of Mira and Intrepid \cite{DT-sc20} sheds light on the evolution of system performance and efficiency over time. 
These studies underscore the importance of log analysis in identifying performance bottlenecks, optimizing resource allocation, and improving efficiency. Unlike these previous studies primarily focused on HPC system performance, this work examines the power efficiency of modern HPC systems, with a particular emphasis on GPU resources.


The collection and analysis of power metrics has long been established on HPC systems, predating the widespread adoption of hybrid CPU-GPU platforms \cite{bridges}. In \cite{DT-ipdps20}, the authors analyze CPU power consumption patterns on two medium-scale European HPC  clusters.
Further efforts to correlate workload data with power metrics have been crucial in enabling power-aware scheduling on HPC systems \cite{7877134}.
And recently, the study of GPU power has gained increasing importance due to the growing prevalence of GPUs and their significant energy costs.
Notable studies include two recent analyses of power patterns in production systems \cite{nersc23,10820672}.
In \cite{nersc23}, the authors investigate the power characteristics of the Perlmutter system at NERSC, highlighting significant power fluctuations and distinct power usage patterns across applications. Their analysis primarily focuses on node-level power consumption through benchmarking, using NERSC-10 workflow benchmarks that represent production workloads. 
In \cite{10820672},  Zhao et al. examine the impact of software-driven power management techniques like DVFS and power capping by leveraging GPU benchmarking and telemetry data.
These studies use a combination of benchmark profiling and system logs to reach their goals.

In contrast, our work adopts a comprehensive approach by jointly analyzing multiple system log sources to characterize system-wide GPU power demand. This integrated analysis yields key insights for improving GPU power efficiency in modern HPC systems.
Furthermore, the system logs and the analysis tool will be released as open-source, providing a practical and reproducible contribution to energy-aware computing.


\section{Conclusions}
\vspace{-1.5mm}

In this work, we presented a systematic approach for co-analyzing diverse telemetry data collected from a production system, uncovering key patterns in workload characterization and their GPU resource usage. 
Through in-depth statistical analysis, we identified notable relationships between workload characteristics and GPU resource demands, including GPU cores, GPU memory, and GPU power.
Given these insights, we then highlighted practical opportunities for applying power savings strategies.   
We anticipate these insights will guide HPC researchers seeking to optimize energy efficiency.  

Future work includes domain analysis to investigate how scientific fields correlate with workload and power characteristics, as well as extending this analysis to additional supercomputing systems.


\bibliographystyle{ACM-Reference-Format}
\bibliography{ref}

\appendix



\end{document}